# Direct X-Ray Measurements of Strain in Monolayer MoS$_2$ from Capping Layers and Geometrical Features

Kathryn Neilson[1], Marc Jaikissoon[1], Dante Zakhidov[2], Tara Peña[1], Alberto Salleo[2], Krishna Saraswat[1,2], and Eric Pop[1,2,3,*]

[1] *Department of Electrical Engineering, Stanford University, Stanford, CA 94305*

[2] *Department of Materials Science and Engineering, Stanford University, Stanford, CA 94305*

[3] *Department of Applied Physics, Stanford University, Stanford, CA 94305*

*Contact: epop@stanford.edu

**ABSTRACT:** Strain induced through fabrication, both by patterning and capping, can be used to change the properties of two-dimensional (2D) materials or other thin films. Here, we explore how capping layers impart strain to monolayer MoS$_2$ using direct x-ray diffraction measurements of the lattice. We first observe the impact of naturally-oxidized metal layers (~1.5 nm Al) and subsequently-deposited Al$_2$O$_3$ (15 nm to 25 nm thick) on the 2D material, and find that the strain imparted to MoS$_2$ is mainly controlled by the interfacial adhesion of the seed layer in addition to the substrate adhesion. Then, using test structures which mimic transistor contacts, we measure enhanced strain from such patterns compared to blanket films. Furthermore, we observe significant tensile strain—up to 2% in monolayer MoS$_2$, one of the largest experimental values to date on a rigid substrate—due to highly-stressed blanket metal capping layers. These results provide direct evidence supporting previous reports of strain effects in 2D material devices.

Atomically thin two-dimensional (2D) materials have gained much attention since monolayer graphene was isolated and electrically probed[1] in 2004. In particular, the sub-nanometer thickness of 2D semiconductor monolayers (the most common being transition metal dichalcogenides, or TMDs) makes them ideal for transistor scaling.[2] When integrating such materials on various substrates, strain is often present, either from the high-temperature synthesis process or from subsequent fabrication steps, contacts, or capping layers.[3-7] Even small amounts of strain (~1%) can be used in a wide variety of applications, including (but not limited to) band gap tuning,[8,9] phase transformations,[10-12] inducing novel quantum states,[13] and in conventional silicon technology.[14] Strain engineering can therefore provide an opportunity to enhance properties of 2D materials by changing their band structure and associated properties.

Recent studies[5,15-20] have found that tensile strain applied to monolayer (1L) TMDs increases the energy separation between the (lower) conduction band K valley and (upper) Q valley, decreasing the intervalley scattering of electrons and increasing their mobility. A similar principle is emerging for the valence band of TMDs, although the performance benefit is reported to arise with compressive strain[21,22] due to the increased valley separation in the valence band.[23] To estimate the strain, reports[24,25] have used indirect optical methods like photoluminescence and Raman spectroscopy. However, the analysis of the data obtained with these techniques can lead to inaccurate strain estimates because the materials' response must be calibrated separately and is also affected by dopants, defects, and the environment.[26] The most *direct* approach is to measure the lattice spacing of the crystal, and hence strain, by x-ray diffraction (XRD), the gold standard in structural materials characterization. An additional benefit of XRD over techniques such as photoluminescence and Raman is that XRD measures a larger area, and that XRD can measure through thicker metal.

In this work, we use XRD to directly measure the strain imparted in monolayer $MoS_2$ from its synthesis (on $SiO_2$/Si), from capping with common uniform layers (e.g. $AlO_x$), and from parallel metal lines which mimic transistor-like geometries. We find that the atomic layer deposition (ALD) temperature[27] and thickness of the $AlO_x$ capping has little impact on the strain imparted to the underlying 2D film. Nevertheless, we observe a small strain (~0.1%), which is mainly due to the seed layer used for ALD nucleation. The strain of the monolayer $MoS_2$ however is increased by ~6 times when Au metal lines capped by $AlO_x$ are used instead, due to the geometry effect on strain distribution. Finally, we use a highly strained blanket capping layer of Au/Ti/Ni on transferred $MoS_2$ and achieve a measured tensile strain of ~2%, demonstrating potential tunability in the band structure of monolayer 2D materials on rigid, rather than flexible, substrates.

Monolayer $MoS_2$ is grown at 750 °C on 90 nm $SiO_2$ on Si substrates using perylene-3,4,9,10-tetracarboxylic acid tetrapotassium salt (PTAS) as growth promoter[28]. Samples are carefully selected for large-area coverage and 50-100 μm grain size [shown in **Fig. 1(a)**]. Most $MoS_2$ films were examined as-grown, but a few were also examined after transfer to a new $SiO_2$/Si substrate. If transferred, $MoS_2$ chips are coated with a polystyrene/toluene mix (3 g/20 mL), delaminating the $MoS_2$ in water, and finally picking up the $MoS_2$/polymer with a target substrate. The polymer is subsequently removed in solvent, and a further anneal (250 °C, 2 hrs, in ~$10^{-6}$ Torr) is implemented to better adhere the $MoS_2$ to the substrate. Then, capping layers were deposited as blanket films onto $MoS_2$ with and without patterning and metallization.

Grazing incidence x-ray diffraction (GIXRD) is carried out on beamline 10-2a at SLAC National Accelerator Lab [**Fig. 1(b)**], using a 14 keV source ($\lambda = 0.8856$ Å). A 3D-printed PET (polyethylene terephthalate) holder clasps samples at a 90° angle through a vacuum attachment at the back of the holder, and Pb foil shields are placed at the front of the holder, nearest to the incident beam, to prevent additional background scattering from the polymer [**Fig. 1(c)**]. Finally, the holder is covered in x-ray transparent thermal tape and filled with $N_2$ ambient gas to reduce noise from air scattering and mitigate beam damage. Incident x-rays enter through the side of the PET holder, diffract off the $MoS_2$ chip, are filtered through Soller slits, and finally absorbed by the detector, as shown in **Fig. 1(d)**.

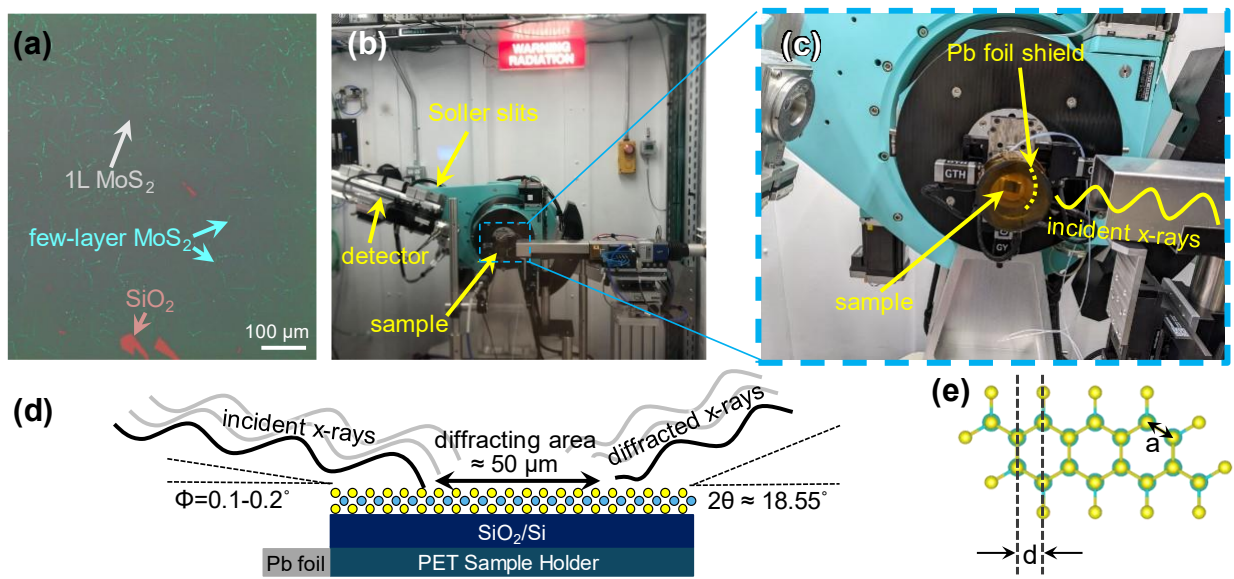

**FIG. 1. (a)** Optical image of coalesced monolayer (1L) MoS$_2$ with few-layer regions along the grain edges. **(b)** Hutch station for grazing incidence x-ray diffraction (GIXRD), where x-rays are incident from the right side of the image, diffract off the sample, are filtered through the Soller slits, and finally absorbed by the detector. **(c)** Close-up image of the GIXRD setup, including a Pb foil shield which helps absorb scattered x-rays from the polymer sample holder. The holder is sealed with thermal tape and filled with N$_2$ gas. **(d)** Schematic of the GIXRD scattering process. The grazing incident angle is 0.1-0.2° (depending on the sample), corresponding to a ~50 μm diffraction spot size. **(e)** Top-down view of monolayer MoS$_2$ crystal structure. The x-ray probes the $d = \sqrt{3}a/2$ spacing, where $a$ is the lattice constant. We measure the in-plane lattice spacing to evaluate strain as relevant for charge transport.

X-rays diffract off the MoS$_2$ lattice according to Bragg's Law, $n\lambda = 2d\sin(\theta)$, where $n$ is a positive integer, $\lambda$ is the incident x-ray wavelength, $d$ is the spacing labeled in **Fig. 1(e)**, and $\theta$ is the diffracted angle. For 2D materials, we measure the (0l) peak to directly evaluate in-plane strain, which is of interest to examine how capping layers[5] and metal contacts[6] may affect planar transistors. Resulting diffraction peak positions and the corresponding $d$ spacing for all spectra in this study are found in **Tables S1-S6**.

First, we measured the as-grown strain in monolayer MoS$_2$ by comparing as-grown materials with films transferred onto a bare 90 nm SiO$_2$ on Si chip. From **Fig. 2(a)**, it is confirmed that the as-grown monolayer MoS$_2$ tends to have built-in tensile strain,[3,29] here ~0.5% for CVD growth at 750 °C. After the layer transfer onto a new SiO$_2$/Si substrate, the built-in strain is released, and some compressive strain may be introduced, depending on the transfer method. Tensile strain in the as-grown MoS$_2$ arises due to a mismatch between the coefficient of thermal expansion (CTE) of the SiO$_2$ substrate and the 2D material during the cool-down process after growth. Transferred MoS$_2$ will have lower adhesion to the substrate, relying on the weak van der Waals interaction of the 2D material and the SiO$_2$ underneath.[4,30]

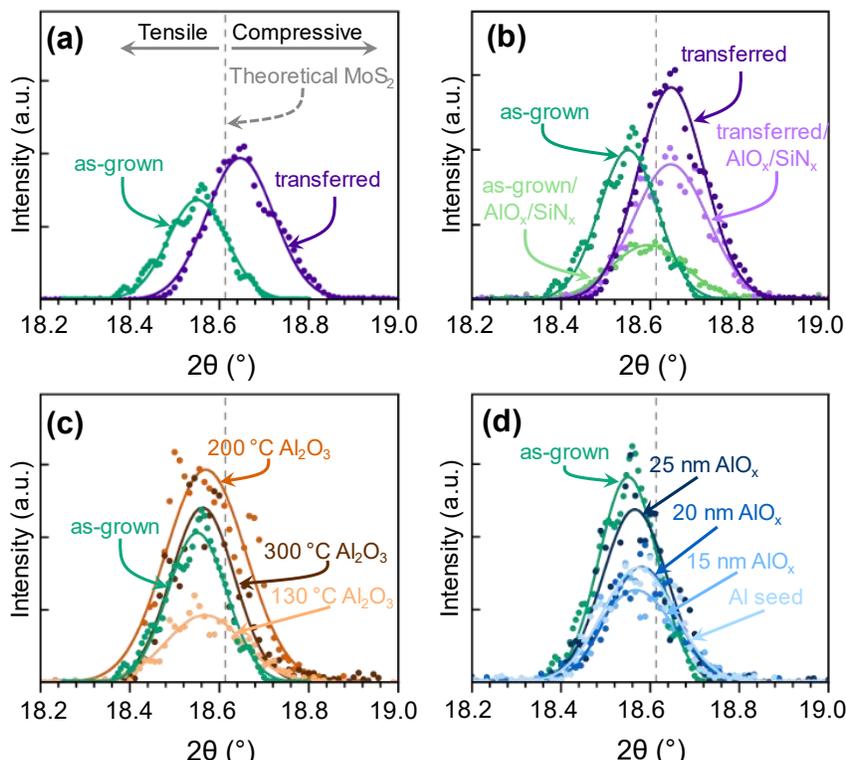

**FIG. 2. (a)** GIXRD comparing as-grown monolayer $MoS_2$ vs. transferred $MoS_2$ onto the same substrate, 90 nm $SiO_2$ on silicon. The built-in tensile strain ($\varepsilon$) from thermal expansion mismatch during growth is released upon transfer. Symbols represent experimental data, lines are Gaussian fits. The change in strain compared to theoretical lattice spacing is $\varepsilon = 0.33\%$ for as-grown, and $-0.18\%$ for transferred $MoS_2$. **(b)** Comparing the strain imparted into $MoS_2$ (using ~1.5 nm Al seed layer, followed by 5 nm $Al_2O_3$ and 15 nm $SiN_x$) on an as-grown $MoS_2$ sample and a transferred $MoS_2$ sample. The transferred sample sees no change in the diffraction angle, while the as-grown sees a shift to less tensile strain, indicating that the adhesion to the substrate plays a critical role in strain transfer. The change in strain with the additional alumina and silicon nitride layers is $\varepsilon = -0.23\%$ for as-grown, and $0\%$ for transferred $MoS_2$. **(c)** Comparing the effect of deposition temperature of ALD alumina (~15 nm thick) on the strain it imparts into monolayer $MoS_2$. All peaks show a similar center position, indicating that the ALD deposition temperature has little impact on strain imparted into $MoS_2$. The change in strain with increasing temperature follows as $\varepsilon = -0.08, -0.09, -0.06\%$. **(d)** Varying the thickness of alumina to test how the ALD encapsulation changes strain in $MoS_2$. All samples start with 1.5 nm Al seed, followed by subsequent ALD at 130 °C; all peaks show a similarly small right-shift to higher angles, as the Al seed at the 2D interface plays the biggest role. The change in strain with increasing thickness follows as $\varepsilon = -0.12, -0.08, -0.15\%$, and $-0.07\%$ for Al seed, 15 nm, 20 nm, and 25 nm of alumina, respectively. All diffraction peaks shown are the (01) peak of $MoS_2$. Details for strain calculations can be found in **Tables S1-S4** for **(a)-(d)**, respectively.

Another experiment is shown in **Fig. 2(b)**, where a capping layer consisting of ~1.5 nm Al seed with 5 nm ALD $Al_2O_3$ and 15 nm $SiN_x$ transfers significant compressive strain ($-0.23\%$) to an as-grown $MoS_2$ film (indicated by the higher angle of the diffraction peak), but the same amount of compressive strain is not imparted in $MoS_2$ transferred onto a new substrate, revealing the importance of substrate adhesion in strain transfer.[31] This has important implications for engineering the strain, as the adhesion to the substrate must also be optimized, in addition to the in-plane stresses to obtain the desired change in band structure.

ALD films are commonly deposited onto 2D materials for a variety of purposes, including doping,[32] capping,[33] and top-gate dielectrics;[34,35] therefore, it is important to understand if, and how, they affect the 2D material strain. Because the ALD process requires attachment to partially filled dangling bonds, seed layers (such Al or Si, which then oxidize[34]) are often deposited in sub-2 nm thickness to provide a template for further oxide film growth. In **Fig. 2(c)** we find that ALD alumina at different temperatures[27] does not lead to different amounts of strain; rather, the imparted strain (-0.1%) is due to the initially evaporated seed layer. Increasing alumina thickness does not impart additional strain either [**Fig. 2(d)**], suggesting that ALD of oxide films may not be the most effective approach for intentionally straining 2D material devices, as amorphous films are more able to easily re-coordinate and relax strain in the system.

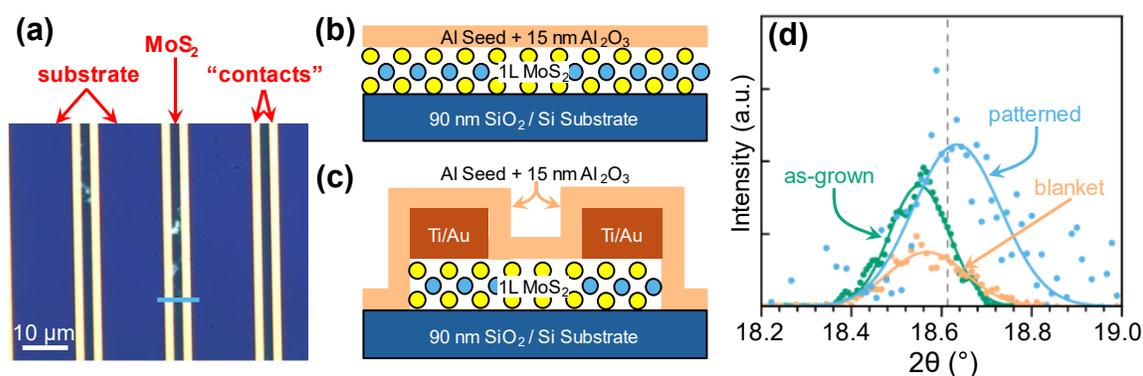

**Fig. 3. (a)** Top-down optical image of the test structure for GIXRD measurements of patterned samples. Long strips of $MoS_2$ with metal contacts on either side (5 nm Ti/ 45 nm Au) serve as an analog to a conventional transistor. The x-ray beam passes parallel to the metal contacts during measurement. The entire sample is covered with Al seed and 15 nm ALD $Al_2O_3$. **(b)** Cross-section schematic of blanket alumina-capped $MoS_2$ films shown in the XRD spectra, and **(c)** patterned alumina-capped $MoS_2$, across the blue line shown in (a). **(d)** Comparing GIXRD spectra of $MoS_2$ films that are as-grown (not capped), blanket-capped as in (b), and patterned then capped as in (a) and (c). The patterned sample sees a greater shift to higher angles, indicating a larger quantity of strain (here, approximately -0.44%) due to geometry as opposed to the blanket films (-0.08%). Calculation and fitting details can be found in supplementary material **Table S5**.

Thus far, we have examined strain imparted into $MoS_2$ by blanket capping layers; however, it is important to study the strain for different geometries of the capping layer, as strain can depend on the dimensions of the film stressor and/or the presence of metal contacts and gates (i.e. transistor topography). In **Fig. 3(a)** we pattern the $MoS_2$ into strips with long metal "contact" lines, to mimic a transistor-like geometry, keeping in mind that individual devices (with sub-micron dimensions) would not yield sufficient GIXRD signal-to-noise. **Figures 3(b)** and **3(c)** display cross-sections of the control sample (with only alumina blanket film) and the patterned transistor-like sample (with metal line contacts and alumina capping). Au is chosen as the contact, as Au metal lines alone impart minimal strain into the 2D material.[26] The GIXRD results are shown in **Fig. 3(d)**, revealing that the samples with patterned contact lines display ~6× greater strain (-0.44% vs. -0.07%) compared to samples with blanket ALD $Al_2O_3$, giving direct evidence that the geometry can further enhance strain transfer in transistors. We note that the patterned sample has a larger full width at half maximum due to non-uniform strain that is known to occur along the device channel.[6,36,37]

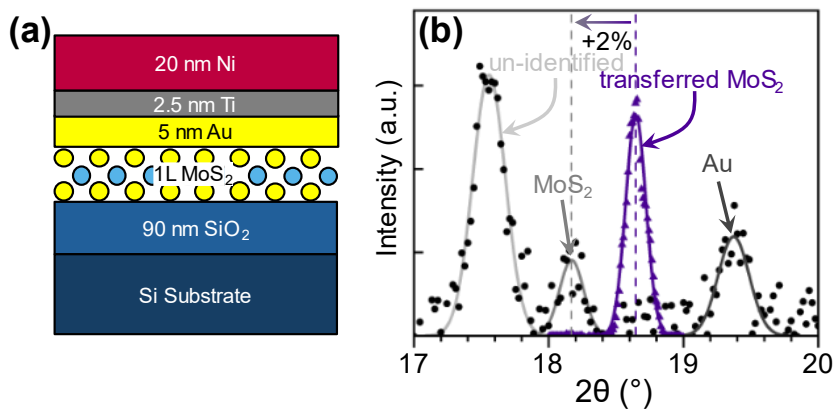

**FIG. 4. (a)** Schematic of a high-strain blanket metal capping layer on monolayer MoS$_2$, similar to the stacks used in Jaikissoon *et al.*[6] Here the MoS$_2$ is transferred, not directly grown, on this SiO$_2$/Si substrate. **(b)** Measured x-ray spectra for the MoS$_2$ (purple symbols, without capping) and after the high-stress capping layer (black symbols and line fits). Three peaks are visible for the metal-capped sample: one from Au, one assigned to the 1H MoS$_2$, and a left-most peak that cannot be assigned. Compared to the uncapped, transferred MoS$_2$, this metal-capped stack shows large tensile strain (>2%) imparted into the 2D material. Parameters for the stress calculation can be found in supplementary material **Table S6**.

To utilize the excess strain that can be imparted through geometry, we have previously shown that materials like Ni can be deposited on the contact region, imparting strain to the MoS$_2$ channel and increasing the short-channel device current.[6] Here we use a similar metal stack, illustrated in **Fig. 4(a)**, to study the strain it can impart as a blanket capping layer on MoS$_2$. The stack consists of monolayer MoS$_2$ capped with 5 nm Au, 2.5 nm Ti as a sticking layer, and 20 nm Ni as stressor. **Figure 4(b)** displays the resulting XRD spectra, showing tensile strain on the order of 2% imparted to the MoS$_2$, which is a substantial amount given that typical reports give values on the range of 0.5-1%. To our knowledge, this is the one of the largest magnitudes of strain to date reported on a rigid substrate imparted into a monolayer 2D material. Factors such as adhesion to the substrate, geometry from patterning, and potential cracking of the material under too high of a strain could all play a role in local strain relaxation. However, there are a few distinctions to our findings, which require additional discussion.

First, most reports of large strains have been estimated with Raman spectroscopy, which gives less accurate results (due to doping or plasmon coupling effects) and is unable to probe through thicker metal layers.[26] And second, we observe a diffraction peak at the smallest angle (~17.6°) in the metal-capped sample. While this peak cannot be confidently attributed to a particular phase (**Table S7** lists possible phases and their corresponding *d* spacing), the closest intermetallic to our knowledge is metastable TiAu, formed by ion irradiation. This peak could also be due to strain non-uniformity, as previously suggested.[37] Alternatively, if large strains are present, there could be localized areas where highly-strained 1T' $MoS_2$ is present. Theoretical analysis of the 1H to 1T' transition[38,39] has suggested that a phase transformation could occur with large amounts of tensile strain, greater than 10% for biaxially strained $MoS_2$. However, factors such as mixed phase forms of 2D materials, phonon vibrations, and salt incorporation (i.e. the potassium from PTAS) from the growth may decrease the amount of strain for a phase transformation.[40,41] We note that contact electrodes on 2D materials with this metal stack (Au/Ti/Ni) are be unlikely to fully benefit from such large strains, because the strain profile in the $MoS_2$ changes rapidly under the contact (within 10s of nm),[36] unlike the blanket case presented here.

In summary, we used grazing-incidence x-ray diffraction to directly measure the strain imparted in monolayer $MoS_2$ by common capping layers ($AlO_x$, $SiN_x$), contact metals (Ni), as blanket layers and in transistor-like contact geometries. We find that strain imparted from $Al_2O_3$ by ALD is controlled by the Al seed layer deposited on the 2D material, and not by the $Al_2O_3$ thickness or ALD temperature. We also found that long metal lines deposited on $MoS_2$, which mimic transistor contacts, can provide ~6 times higher strain compared to blanket $Al_2O_3$ layers. Blanket metal stacks with stressed Ni can impart even greater strain in $MoS_2$, up to 2%. Further data analysis of diffraction line shapes may provide more information on the uniformity of the strain. These x-ray measurements directly probe the (average) lattice spacing, providing valuable insight into tuning the band structure of monolayer 2D materials using insulating and metal capping layers.

This work was performed in part at the Stanford Nanofabrication Facility (SNF) and Stanford Nano Shared Facilities (SNSF), RRID:SCR_023230, supported by the National Science Foundation (NSF) under Award ECCS-2026822. K.N. acknowledges the Stanford Graduate Fellowship (SGF) program, and K.N. and D.Z. acknowledges the NSF Graduate Research Fellowship under Grant No. DGE-1656518. Synchrotron measurements were done at the SLAC National Accelerator Laboratory, supported by the Department of Energy under Contract No. DE-AC02-76-SF0015. The work was supported in part by the Stanford SystemX Alliance. K.N. and E.P. acknowledge partial support by Intel Corp. and NSF FuSe2 Award 2425218, M.J. and K.S. acknowledge the Samsung Global Research Outreach (GRO) program.

## AUTHOR DECLARATIONS

### Conflict of Interest

The authors have no conflicts to disclose.

### Acknowledgements

The authors would like to thank Dr. Christopher Takacs from SLAC for his assistance on the beamline.

### Contributions

K.N. and D.Z. performed the GIXRD measurements. M.J. synthesized the MoS$_2$. K.N. did associated patterning of MoS$_2$ samples, and K.N. and M.J. deposited films onto MoS$_2$. All authors assisted with the preparation of the manuscript. A.S., K.S., and E.P. supervised the project.

## DATA AVAILABILITY

The data supporting this study are available from the corresponding author upon reasonable request.

# Supplementary Material

# Direct X-Ray Measurements of Strain in Monolayer $MoS_2$ from Capping Layers and Geometrical Features

Kathryn Neilson[1], Marc Jaikissoon[1], Dante Zakhidov[2], Tara Peña[1], Alberto Salleo[2], Krishna Saraswat[1,2], and Eric Pop[1,2,3] *

[1] *Department of Electrical Engineering, Stanford University, Stanford, CA 94305*

[2] *Department of Materials Science and Engineering, Stanford University, Stanford, CA 94305*

[3] *Department of Applied Physics, Stanford University, Stanford, CA 94305*

*Contact: epop@stanford.edu

## S1. Strain calculation details

X-ray diffraction spectra are fit using a Gaussian curve, centered at $2\theta$, representing an averaged value over the x-ray spot size. The tables below reports the films probed, their corresponding lattice spacing, and strain. $MoS_2$ films are measured as-grown (at 750 °C on $SiO_2$/Si), unless described as "transferred" (onto a new $SiO_2$/Si substrate). Strain percentages are calculated with respect to their preparation method, i.e. as-grown or transferred $MoS_2$. As-grown and transferred $MoS_2$ with no encapsulation layer are calculated in reference to the theoretical value. Negative strains indicate imparted compressive strain.

**Table S1**: Comparing as-grown and transferred $MoS_2$ to expected $MoS_2$ lattice spacing.

| Film | $2\theta$ (°) | $d$ (Å) | Strain (%) |
|---|---|---|---|
| Expected $MoS_2$[1,2] | 18.613 | 2.738 | |
| As-grown $MoS_2$ | 18.552 | 2.747 | 0.33 |
| Transferred $MoS_2$ | 18.646 | 2.733 | -0.18 |

**Table S2**: Comparing imparted strain on as-grown versus transferred $MoS_2$ films.

| Film | $2\theta$ (°) | $d$ (Å) | Strain (%) |
|---|---|---|---|
| As-grown $MoS_2$ | 18.552 | 2.747 | |
| Transferred $MoS_2$ | 18.646 | 2.733 | |
| $MoS_2$ with Al seed + 5 nm $Al_2O_3$ at 130 °C + 15 nm $SiN_x$ | 18.595 | 2.741 | -0.23 |
| Transferred $MoS_2$ with Al seed + 5 nm 130 °C $Al_2O_3$ + 15 nm $SiN_x$ | 18.646 | 2.733 | 0.00 |

**Table S3**: Comparing the imparted strain on alumina deposited at different temperatures.

| Film | 2θ (°) | d (Å) | Strain (%) |
|---|---|---|---|
| As-grown MoS$_2$ | 18.552 | 2.747 | |
| MoS$_2$ with Al seed + 15 nm Al$_2$O$_3$ at 130 °C | 18.567 | 2.745 | -0.08 |
| MoS$_2$ with Al seed + 15 nm Al$_2$O$_3$ at 200 °C | 18.570 | 2.745 | -0.09 |
| MoS$_2$ with Al seed + 15 nm Al$_2$O$_3$ at 300 °C | 18.563 | 2.746 | -0.06 |

**Table S4**: Comparing the effect of increasing alumina thickness on imparted strain.

| Film | 2θ (°) | d (Å) | Strain (%) |
|---|---|---|---|
| As-grown MoS$_2$ | 18.552 | 2.747 | |
| MoS$_2$ with 1.5 nm Al seed layer | 18.574 | 2.744 | -0.12 |
| MoS$_2$ with 1.5 nm Al seed + 15 nm Al$_2$O$_3$ at 130 °C | 18.567 | 2.745 | -0.08 |
| MoS$_2$ with 1.5 nm Al seed + 20 nm Al$_2$O$_3$ at 130 °C | 18.581 | 2.743 | -0.15 |
| MoS$_2$ with 1.5 nm Al seed + 25 nm Al$_2$O$_3$ at 130 °C | 18.565 | 2.745 | -0.07 |

**Table S5**: Comparing blanket versus patterning films and the effect of patterning on imparted strain.

| Film | 2θ (°) | d (Å) | Strain (%) |
|---|---|---|---|
| As-grown MoS$_2$ | 18.552 | 2.747 | |
| MoS$_2$ with 1.5 nm Al seed + 15 nm Al$_2$O$_3$ at 130 °C | 18.567 | 2.745 | -0.08 |
| MoS$_2$ patterned with Al seed & 15 nm ALD Al$_2$O$_3$ at 130 °C | 18.634 | 2.735 | -0.44 |

**Table S6**: Evaluating the strain in a highly strained film stack with transferred MoS$_2$.

| Film | 2θ (°) | d (Å) | Strain (%) |
|---|---|---|---|
| Transferred MoS$_2$ | 18.646 | 2.733 | |
| Transferred MoS$_2$ with 5 nm Au, 2.5 nm Ti, and 20 nm Ni | 18.170 | 2.80 | 2.45 |

**Table S7**: Metal diffraction lattice spacings corresponding to the structure shown in **Fig. 4a**. The choice of plane and subsequent *d* spacing for each material is based on those closest to the measured $2\theta$ range and the theoretical intensity of the plane.

| Material | Bravais Lattice | Space Group | d (Å) | Plane | Reference |
| --- | --- | --- | --- | --- | --- |
| Au | Cubic | Fm-3m | 2.355 | (111) | 3 |
| Ti | Hexagonal | P63/mmc | 2.555 | (100) | 4 |
| Ni | Cubic | Fm-3m | 2.034 | (111) | 3 |
| $Au_2Ti$ | Tetragonal | I4/mmm | 2.425 | (110) | 5 |
| $Au_4Ti$ | Tetragonal | I4/m | 2.337 | (121) | 6 |
| $AuTi_3$ | Cubic | Pm-3n | 2.549 | (200) | 7 |
| AuTi | Hexagonal | P63/mmc | 2.832 | (100) | 8 |
| AuTi | Cubic | Pm-3m | 3.250 | (100) | 9 |
| $Au_3Ti$ | Cubic | Pm-3n | 2.550 | (200) | 9 |
| $Au_{0.36}TiNi_{0.64}$ | Cubic | Pm-3m | 3.087 | (100) | 10 |
| $Au_{0.36}TiNi_{0.64}$ | Orthorhombic | Pmma | 3.190 | (101) | 10 |
| $Au_3Ni$ | Cubic | Pm-3m | 2.828 | (110) | 11 |
| AuNi | Cubic | Fm-3m | 2.218 | (111) | 12 |
| $Ni_3Ti$ | Hexagonal | P63/mmc | 2.132 | (201) | 13 |
| $Ni_4Ti_3$ | Hexagonal | R-3 | 2.128 | (410) | 14 |
| NiTi | Hexagonal | P3 | 2.138 | (11-2) | 15 |
| $NiTi_2$ | Cubic | Fd-3m | 2.302 | (422) | 16 |
| $TiS_2$ | Hexagonal | P-3m1 | 2.622 | (011) | 17 |
| $TiS_3$ | Monoclinic | P21/m | 2.902 | (003) | 18 |
| $MoS_2$ | Hexagonal | P63/mmc | 2.738 | (100) | 1 |
| $MoS_2$ | Hexagonal | P-3m1 | 2.780 | (100) | 19 |
| Fig. 4b, Leftmost Peak | | | 2.894 | | This work |